\documentclass[twocolumn,preprintnumbers,amssymb,amsmath,superscriptaddress,letterpaper]{revtex4}[12pt]

\usepackage{graphicx}
\usepackage{dcolumn}
\usepackage{bm}
\usepackage{natbib}
\usepackage{pstricks}
\usepackage{epsfig}
\usepackage{epstopdf}
\usepackage{epigraph}
\usepackage{mathtools}

\newcommand{\be}{\begin{equation}}
\newcommand{\ee}{\end{equation}}
\newcommand{\bea}{\begin{eqnarray}}
\newcommand{\eea}{\end{eqnarray}}

\begin{document}

\title{\"Uber-Gravity and the Cosmological Constant Problem}

\author{Nima Khosravi}
\email{n-khosravi@sbu.ac.ir}
\affiliation{Department of Physics, Shahid Beheshti University, G.C., Evin, Tehran 19839, Iran}

\date{\today}

\begin{abstract}
Recently, the idea of taking ensemble average over gravity models has been introduced. Based on this idea, we study the ensemble average over (effectively) all the gravity models (constructed from Ricci scalar) dubbing the name \"uber-gravity which is a {\it{fixed point}} in the model space. The \"uber-gravity has interesting universal properties, independent from the choice of basis: $i)$ it mimics Einstein-Hilbert gravity for high-curvature regime, $ii)$ it predicts stronger gravitational force for an intermediate-curvature regime, $iii)$ surprisingly, for low-curvature regime, i.e. $R<R_0$ where $R$ is Ricci scalar and $R_0$ is a given scale, the Lagrangian vanishes automatically and $iiii)$ there is a sharp transition between low- and intermediate-curvature regimes at $R=R_0$. We show that the \"uber-gravity response is robust to all values of vacuum energy, $\rho_{vac}$ when there is no other matter. So as a toy model, \"uber-gravity, gives a way to think about the hierarchy problems e.g. the cosmological constant problem. Due to the transition at $R=R_0$ there is a chance for \"uber-gravity to bypass Weinberg's no-go theorem. The cosmology of this model is also promising because of its non-trivial predictions for small curvature scales in comparison to $\Lambda$CDM model.

\end{abstract}

\maketitle

\section{Introduction:}
A century ago Einstein introduced the cosmological constant (CC) to address static universe \cite{CC-Einstein-1917} which became his biggest blunder after Hubble's discovery of expanding universe. On the other hand, from the viewpoint of particle physics it is well-known that there is a non-vanishing vacuum energy, $\rho_{vac}$, which has no effect on most of particle physics' calculations. But in presence of gravity, it predicts an inflating universe which is not compatible with the observations before 1998. Accordingly it raised a question: why the vacuum energy has no effect on gravity? which is known as old CC-problem. Data acquired by Sueprnovae observations in 1998 \cite{acceleration} and recent Plank data \cite{planck} implies a tiny value for CC, which shall be 120 orders of magnitude smaller than $\rho_{vac}$; this prediction sometimes will refer to as ``the worst theoretical prediction in the history of physics" \cite{hobson}. To solve this discrepancy a fine-tuning is required which is known as the new CC-problem (CCP) \cite{CC-problem}.

There are three different approaches to solve the CCP: $i)$ modifying the Einstein-Hilbert (EH) model in a way that gravity becomes insensitive to $\rho_{vac}$ \cite{CC-MG}, $ii)$ revising field theory calculation of $\rho_{vac}$ \cite{CC-FT} and $iii)$ connecting the CCP (which is in IR regime) to UV-completion of gravity \cite{CC-QG}. An idea in the context of modifying gravity is degravitation  which proposes switching off the gravity for very large wavelengths and consequently filters $\rho_{vac}$ \cite{degravitation}. It is worth mentioning that some believes old CCP shall be addressed before moving to new CCP. This idea is supported by 'tHooft conjecture: if the gravitational effects of $\rho_{vac}$ can be canceled by a symmetry then a tiny fluctuation from this symmetric situation is natural. Supersymmetry is an idea in this direction  assuming the presence of a boson particle for each fermion consequently paving the way for a mechanism to eliminate $\rho_{vac}$ \cite{supersym}.

In this paper, we will study the CCP within the context of \"uber-modeling introduced in \cite{uber}. We try to show that \"uber-modeling of (effectively) all gravitational models eliminates gravity for low-curvature regimes which can be interpreted as degravitation. Our model coincides with the EH model in high-curvature regime although, there is an intermediate-curvature regime where gravity is stronger than the standard  EH model. It will be shown that our model is not sensitive to  value of vacuum energy, $\rho_{vac}$, thanks to a sharp (but continuous) transition from low- to intermediate-curvature regime. Interestingly, this means there is no need of fine-tuning and the CC is ``natural"\footnote{This result has been shown for a certain circumstance where all the matter fields are turned off except the vacuum energy. At this step our model is same as the first steps in other models e.g. unimodular scenario and should be studied for more details in the future.}. 
\\

\section{\"Uber-gravity:}
\begin{figure}
	\centering
	\includegraphics[width=1\linewidth]{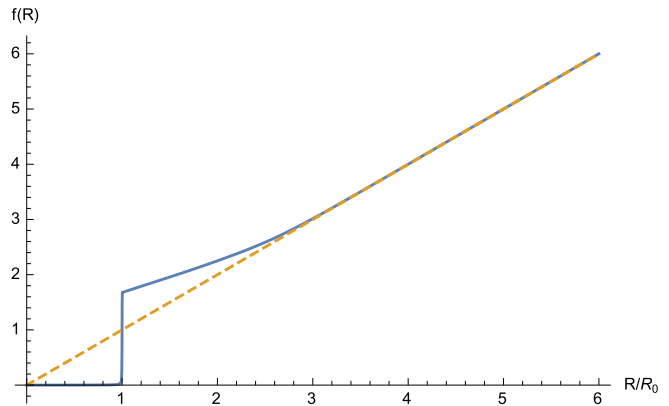}
	\caption{Blue line is our Lagrangian (\ref{uberfR1}) where we do sum up to $N=1000$ (It is easy to see that for larger $N$'s the plot is practically the same.) and yellow dashed line shows the EH action for comparison. The universal behavior of our model is obvious: i) in high-curvature regime our model coincides with the EH model, ii) in intermediate-curvature regime where gravity is stronger than the EH model, iii) for $R<R_0$ gravity vanishes and iiii) there is a sharp transition at $R=R_0$.}
	\label{fig:fR1}
\end{figure}

\begin{figure}
	\centering
	\includegraphics[width=1\linewidth]{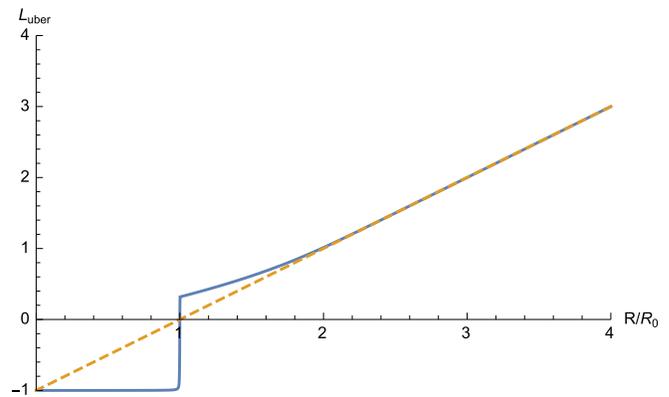}
	\caption{We plotted the \"uber-gravity Lagrangian with $R^n-\lambda_n$ as our basis. We assumed $\lambda_n=R_0^n$ and obviously our model mimics GR plus cosmological constant for high-curvature regime.}
	\label{fig:fR-lambda}
\end{figure}

\begin{figure}
	\centering
	\includegraphics[width=1\linewidth]{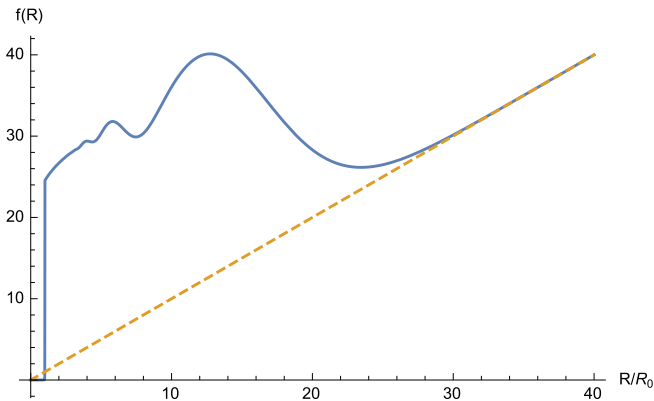}
	\caption{Blue line shows our Lagrangian (\ref{uberfR1}) where $\beta=0.01$ while yellow dashed line represents the EH action for comparison. This plot demonstrates that our model has  non-trivial features in its intermediate-curvature regime depending on the value of $\beta$. However we emphasize that the universal features are the same as $\beta=1$ case, see FIG. \ref{fig:fR1}. However we are not interested in $\beta<1.3$ since they may have instabilities.}
	\label{fig:beta=0.01}
\end{figure}

In \cite{uber}, we introduced an idea based on ensemble average of models within the context of gravity. According to this idea, we start with the space of all  consistent models of gravity, $\mathbb M$, and then take an ensemble average over all models. This idea is inspired by statistical mechanics which employed in a very different context. In \cite{nnaturalness}, Arkani-Hamed et al. employed a similar idea to address the hierarchy problem in particle physics. They mention that in principle an average should be taken on all  possible models but for simplicity, they just considered the standard model with different Higgs masses. The main idea behind addressing the hierarchy problem in both \cite{uber} and \cite{nnaturalness} is a dynamical mechanism which can make our current model dominant. In \cite{nnaturalness} this mechanism is realized by introducing a new field, named reheaton, which ``deposits a majority of the total energy density into the lightest sector" (which is our observed standard model of particle physics). In our \"uber-modeling this mechanism is given by the assigned probability to each model which is introduced by hand at this step. On the other hand our idea can be seen as a realization of the Tegmark's mathematical universe idea \cite{tegmark}, specially when he argues ``all logically acceptable worlds exist". In \cite{uber} we assumed all the theoretically possible (gravity) models play a role in the final model (of gravity). To make \"uber-modelling idea applicable, we assigned a Lagrangian to each model and define (ensemble) average of all the Lagrangians as following:
\begin{eqnarray}\label{lagrangian-general}
{\cal L}= \bigg(\displaystyle\sum_{i=1}^{N}{\cal L}_i e^{- \beta{\cal L}_i}\bigg) \bigg/ \bigg(\displaystyle\sum_{i=1}^{N} e^{- \beta{\cal L}_i}\bigg),
\end{eqnarray}
where $\beta$ is a free parameter and model space is represented by $\mathbb M = \{{\cal L}_i\mid  i\in \{1,N\}\}$ while $N$ is  number of all possible models. We emphasize that the above formulation is inspired by ensemble average procedure in statistical mechanics. However our suggested probabilities are fundamentally different with what is in statistical mechanics. As it is obvious from (\ref{lagrangian-general}) that we use the Lagrangian in the exponent while in statistical mechanics it is $E_i$ which is energy of each state.  The above Lagrangian can be beautifully re-written in a more compact form as
\begin{eqnarray}\label{model}
{\cal L}=-\frac{d }{d \beta}\,ln{\cal Z}, \hspace{2cm}{\cal Z}=\sum_{n=1}^{N}\, e^{-\beta {\cal L}_n}
\end{eqnarray}
which reminds us of the partition function and its relation to energy. In \cite{uber} we assumed  $\mathbb M =\{R,G\}$ where $R$ is the Ricci scalar and $G$ is the Gauss-Bonnet term. In this paper we generalize the model space to (effectively) all the gravity models based on curvature scalar: all analytic $f(R)$. Schematically we can write corresponding partition function as 
\begin{eqnarray}\label{model-fr}
{\cal Z}=\sum_{f(R)}\, e^{-\beta f(R)}.
\end{eqnarray}
Here we deal with analytic functions of $f(R)$ and we can arbitrarily choose the basis. We are working with $\mathbb M=\{R^n \mid \forall\, n\in \mathbb N\}$. The ensemble averaged Lagrangian takes the following form
\begin{eqnarray}\label{uberfR1}
{\cal L}= \bigg(\displaystyle\sum_{n=1}^{\infty}\bar{R}^n e^{- \beta \bar{R}^n}\bigg) \bigg/ \bigg(\displaystyle\sum_{n=1}^{\infty} e^{- \beta \bar{R}^n}\bigg),
\end{eqnarray}
where $\bar{R}=R/R_0$. This model, which belongs to $f(R)$ family, has two free parameters: $R_0$ and dimensionless $\beta$. In FIG. \ref{fig:fR1}, the above Lagrangian is plotted for $\beta=1$.  The above Lagrangian belongs to the $f(R)$ family and effectively is ensemble average of all  possible models of gravity  based on the curvature tensor. There is a possibility to add a constant to each $f(R)$, i.e. working with $R^n-\lambda_n$ as our basis. We plotted its Lagrangian in FIG. \ref{fig:fR-lambda} for $\lambda_n=\lambda$, which mimics GR plus a cosmological constant. This model with additional constant has been studied in \cite{Khosravi:2017hfi} with very interesting observational consequences. But in this work we focus on (\ref{uberfR1}) to study the theoretical properties of the model. Note that in a general case we could work with all  possible linear combinations e.g. ${\cal L}= \alpha_1 R+\alpha_2 R^4$ with two constants $\alpha_1$ and $\alpha_2$. It is easy to observe that adding such terms do not change the interesting aspects of the model, so without loss of generality we focus on the above Lagrangian. Only difference will be in the form of the Lagrangian over the intermediate-curvature regime while for both high- and low-curvature regimes nothing is changed. Even in the intermediate-curvature regime the general prediction is a stronger gravity compared to the EH model. 
However, the form of our model in the intermediate-curvature regime is sensitive to the parameter $\beta$. As an example FIG.\ref{fig:beta=0.01} shows the Lagrangian (\ref{uberfR1}) for $\beta=0.01$ which represents a very different behavior in intermediate-curvature regime. 

In summary, the \"uber-gravity model has the following universal properties (independent to the choice of basis i.e. $\mathbb M$):

\begin{itemize}
	\item for high-curvature regime it reduces to the EH action,
	\item for intermediate-curvature regime it predicts a stronger gravity than the EH model, 
	\item it is vanishing  for low-curvature regime ($R<R_0$),
	\item there is a sharp transition at $R_0$.
\end{itemize}

It is worth mentioning that adding the \"uber-gravity (\ref{uberfR1}) to $\mathbb M$ and re-employing the \"uber-modelling procedure cannot affect above universal features \footnote{We thank S. Baghram for pointing out this issue.}. This is a very significant property since it means \"uber-gravity is a fixed point the model space of $f(R)$ models and this makes it remarkable.  In addition above properties are shared for all values of $\beta$. 

\begin{figure}
	\centering
	\includegraphics[width=1\linewidth]{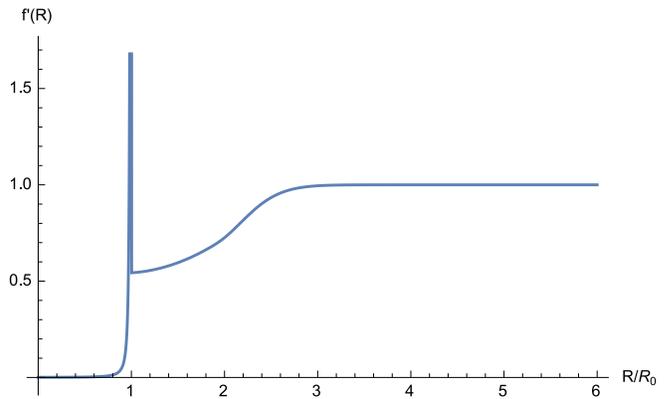}
	\caption{$f'(R)$ is shown for our model (\ref{uberfR1}) where $\beta=1$ though this is not sensitive to $\beta$'s value too much.}
	\label{fig:dfr}
\end{figure}

\begin{figure}
	\centering
	\includegraphics[width=1\linewidth]{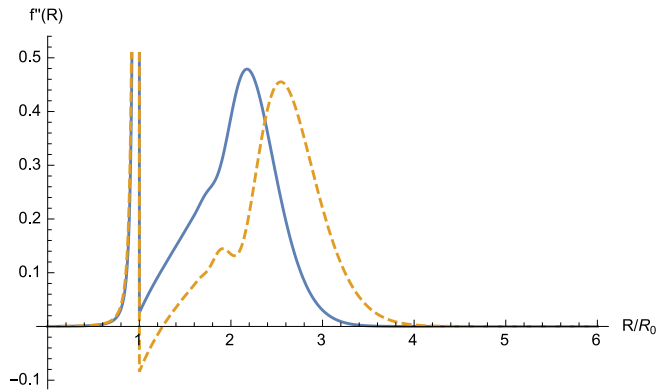}
	\caption{$f''(R)$ is shown for $\beta=1.5$ and $\beta=1$ in solid blue line and yellow dashed line respectively. It is obvious that satisfaction of stability condition, i.e. $f''(R)>0$, depends on the value of $\beta$. For $\beta>1.3$, $f''(R)$ is always positive for $R>R_0$. However in the presence of the matter field we expect to have $R_t>R_0$ where $R_0$ plays the role of the late time cosmological constant. So one can imagine that our model should satisfy the stability conditions even for $\beta\sim 1$. }
	\label{fig:ddfr}
\end{figure}

It is crucial to discuss about the stability of our model (\ref{uberfR1}) which belongs to $f(R)$ models. A dark energy $f(R)$ model is viable if it satisfies $f'(R)>0$ and $f''(R)>0$ for $R\geq R_T>0$ where $R_T$ is the today value of Ricci scalar \cite{fR-review}. In our scenario, $R_0$ will be the late time cosmological constant, so $R_T \rightarrow R_0^+$  in the presence of matter fields. So to show the stability of our model (\ref{uberfR1}) we need to show that $f'(R)>0$ and $f''(R)>0$ for $R>R_0$. It is obvious from FIG. \ref{fig:dfr} and FIG. \ref{fig:ddfr} that there is always an $R>R_0$ where stability conditions can be satisfied. The interesting point is that for $\beta>1.3$, $f''(R)>0$ for $R>R_0$ (see solid line in FIG. \ref{fig:ddfr}) otherwise the condition is satisfied for a larger value than $R_0$. It is important to mention that the sharp transition at $R= R_0$ may behaves like a discontinuity which should be take into account very seriously.

\section{\"Ubergravity and the CCP:}
In this section we study properties of \"uber-gravity model (\ref{uberfR1}). We will show this model is not sensitive to the value of the vacuum energy under a specific circumstances.
\subsection{Equations of Motion:}
For a general $f(R)$ model the equation of motion is as follow \cite{fR-review}:
\begin{eqnarray}\label{eq-motion-fR-matter}\nonumber
\Sigma_{\mu\nu}=\kappa^2 T_{\mu\nu}
\end{eqnarray}
where
\begin{eqnarray}\label{eq-motion-fR}\nonumber
\Sigma_{\mu\nu}=F(R)R_{\mu\nu}-\frac{1}{2}f(R)\,g_{\mu\nu}
+ \bigg(g_{\mu\nu} \Box -\nabla_\mu \nabla_\nu \bigg)F(R),
\end{eqnarray}
$F(R)=\frac{\partial f}{\partial R}$ and $T_{\mu\nu}$ is the energy-momentum tensor. In our case $F(R)$ can be written as:
\begin{eqnarray}\nonumber
F(R)&=&\frac{\left(\displaystyle\sum _{n=1}^\infty R^n e^{-\beta  R^n}\right) \displaystyle\sum
	_{n=1}^\infty \beta  n R^{n-1} e^{-\beta R^n}}{\left(\displaystyle\sum
	_{n=1}^\infty e^{-\beta R^n}\right)^2}\\\nonumber
&-&\frac{\displaystyle\sum _{n=1}^\infty \left(\beta  n
	R^{2 n-1} e^{-\beta R^n}-n R^{n-1} e^{-\beta R^n}\right)}{\displaystyle\sum _{n=1}^\infty e^{-\beta R^n}}.
\end{eqnarray}
It is obvious from the above relations, our model is very complicated for analytical calculations. In the next section we introduce a simplified model which shares all the interesting properties of our model. 
\\

\subsection{Simplified Model:}
The following model has the same features in all the curvature regimes \footnote{We would like to thank Alberto Nicolis because this idea emerged after a discussion with him at the CERN.}
\begin{eqnarray}
f(R)= \begin{cases} \label{simplified-model}
\bar{R}^n & R\leq R_0 \\
\bar{R} +e^{-(\bar{R}-0.7)} & R_0< R 
\end{cases}
\end{eqnarray}
where for the limit $n\rightarrow\infty$ the low-curvature regime shares exactly the same feature with our model, see  FIG.\ref{fig:fR1}. Note that here the exponential term is added phenomenologically in $R_0 < R$ region to recover the intermediate-curvature behavior which mimics (\ref{uberfR1}) for $\beta=1$ and $\lambda_n=0$. In practice by changing $\beta$ one needs to re-calculate parameters in the exponent in simplified Lagrangian (\ref{simplified-model}). For our purpose, we focus on low-curvature regime. It is easy to see that the equation of motion for the low-curvature part is \cite{Rn}:
\begin{eqnarray}\label{sigma-Rn}
&&\Sigma_{\mu\nu}=\\\nonumber&& n\,\bar{R}^{n-1}\bar{R}_{\mu\nu}-\frac{1}{2}\bar{R}^n\,g_{\mu\nu}+n\,R_0^{-1}\bigg(g_{\mu\nu} \Box -\nabla_\mu \nabla_\nu \bigg)\bar{R}^{n-1}
\end{eqnarray}
where $\bar{R}_{\mu\nu}=R_0^{-1}R_{\mu\nu}$. Obviously for high-curvature regime the model reduces to Einstein's gravity. In the next section based on this model we will show how this model can give us a proposal to resolve the CCP.
\\

\subsection{An attempt to solve the CCP:}
To address the CCP  we need to take care of the vacuum energy, $\rho_{vac}$, in presence of gravity. To do this we need to recall that $\rho_{vac}$ is encoded in the trace of energy-momentum tensor, $T$. By looking at (\ref{sigma-Rn}) it is easy to see that the trace of equations of motion yields: 
\begin{eqnarray}
(n-2)\bar{R}^n+3\,n\,R_0^{-1}\,\Box\,\bar{R}^{n-1}=\kappa^2 T.
\end{eqnarray}
We are interested in solutions like $R=cte$ since  for our purpose $T$ is vacuum expectation value which is a constant. With this assumption the above equation reduces to
\begin{eqnarray}\label{trace-rn}
\frac{R}{R_0}=\bigg(\frac{\kappa^2\, T}{n-2}\bigg)^{\frac{1}{n}}.
\end{eqnarray}
For all  $T\neq 0$; limits of equation (\ref{trace-rn}) for $n\rightarrow\infty$ results in $R\rightarrow R_0$. This is a very interesting result which means the model's response to the vacuum energy is robust i.e. gravity sector is not sensitive to $\rho_{vac}$. In other words, the cosmological constant value $R_0$, shall be fixed only by  observation without fine-tuning. Such will imply that the cosmological constant value is natural and the CCP can be solved by this approach. More interestingly, this result is not valid for zero vacuum energy which means particle physics' prediction for non-zero vacuum energy is crucial for our model.
\\

\subsection{A subtlety:}
\begin{figure}
	\centering
	\includegraphics[width=1\linewidth]{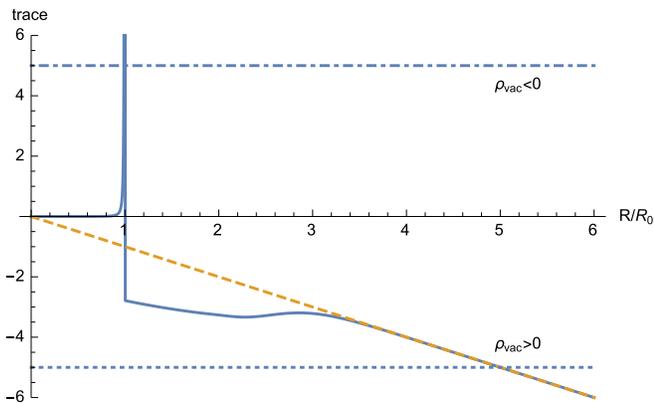}
	\caption{Blue line is the trace of equation of motion in \"uber-gravity where $\beta=1$ and yellow dashed line shows the same for the EH action. Obviously at $R=R_0$ the trace goes asymptotically to positive infinity. It is clear that for positive $\rho_{vac}$ there is only the EH's solution while for any $\rho_{vac}<0$, $R=R_0$ is the solution.}
	\label{fig:trace-full}
\end{figure}

Above argument contains a subtlety which we shall  clarify  herein. The point is that by definition $T\propto - \rho_{vac}$ where $\rho_{vac}>0$; the negative sign is the origin of problem. In the EH model, trace of equation of motion gives $- R = \kappa^2 T$ hence $R=\kappa^2 \rho_{vac}$. But in our scenario (\ref{trace-rn}) for any $n>2$ we have $(-\rho_{vac})^{1/n}$ on the right hand side of equation. Now the question is what is value of $(-1)^{1/n}$ for $n\rightarrow\infty$? For any given $n$ there are $n$ solutions in complex plane and none of them is exactly one. For sure there is a solution which is as close as possible to one but the infinitesimal difference always has an imaginary part. This behavior is shared between both simplified model (\ref{simplified-model}) and the \"uber-gravity (\ref{model}). To illustrate this fact, we plot the trace of equations of motion for \"uber-gravity in FIG. \ref{fig:trace-full}.
We believe that this issue can be addressed by full analysis of the analytical continuation of our model but its concrete study remains open for future investigations. Hereby we will try to give some ideas which can resolve this problem and hopefully guide us to a concrete proposal to solve the CCP.
\\

\subsection{Towards a Proposal to Solve the CCP:}
Here we will try to give two proposals to resolve the above subtlety:

I: Particle Physics approach: The negative sign in $T=-\rho_{vac}$ in standard particle physics is because of larger number of fermionic degrees of freedom compared to bosonic counterpart.  String theory predicts new species like axions \cite{axion}, which are candidates for dark matter \cite{witten}. The axions are bosons which means if one calculates the axion's contribution to the vacuum energy then $\rho_{vac}$ can be negative. As it is obvious from FIG. \ref{fig:trace-full} for a negative $\rho_{vac}$ we have $R=R_0$ as a solution. As mentioned above, this solution is not sensitive to any value of  $\rho_{vac}$ which means there is no need to fine-tuning.

\begin{figure}
		\centering
		\includegraphics[width=1\linewidth]{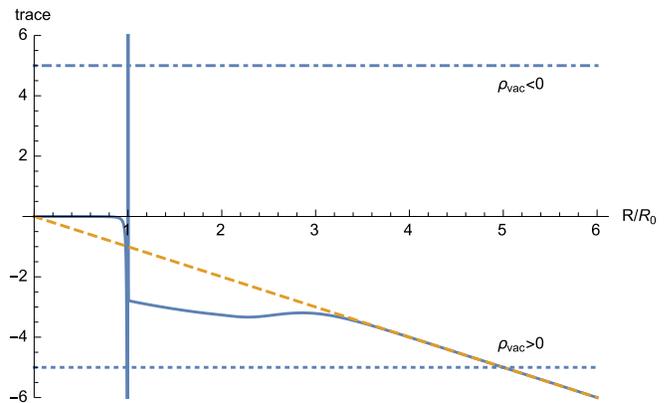}
		\caption{Blue line is the trace of equation of motion in the modified model (\ref{modified-model}) where $\beta=1$ and yellow dashed line shows the same for the EH action. Obviously, at $R=R_0$ the trace goes to negative infinity and comes up to positive infinity asymptotically. In this scenario for  $\rho_{vac}>0$ we get two solutions: $R=R_0$ and $R=\rho_{vac}$. The $R=\rho_{vac}$ solution is same as the EH solution which is not in agreement with the observations. But, the solution $R=R_0$ is the one which can be compatible with observations. This solution is not sensitive to the value of $\rho_{vac}$ which means there is no need to fine-tuning and the CC is natural.}
		\label{fig:trace-full-modified}
	\end{figure}

II: \"Uber-Gravity approach: This approach suggests to modify the gravity model. As it is clear from FIG. \ref{fig:trace-full} what we need is  an asymptotic behavior with an opposite sign at $R=R_0$.  For this purpose, we modify our model (\ref{model}) phenomenologically by multiplying it by a hyperbolic tangent function:
	\begin{eqnarray}\label{modified-model}
	{\cal L}= \tanh\bigg[N (\bar{R}-1)\bigg]\,\,\times\,\, \frac{\displaystyle\sum_{n=1}^{N}\bar{R}^n e^{- \beta \bar{R}^n}}{\displaystyle\sum_{n=1}^{N} e^{- \beta \bar{R}^n}}.
	\end{eqnarray}
Obviously $\tanh$ function changes its sign at $R=R_0$ and behaves as a step function  while $N\rightarrow\infty$. Using $\tanh$ function instead of step function makes our Lagrangian continuous which is useful for future purposes. For this model the trace of equation of motion is plotted in FIG. \ref{fig:trace-full-modified}. In this scenario for a positive $\rho_{vac}$ there are two distinguishable solutions $R=T$ and $R=R_0$. The $R=T$ solution is exactly the EH solution which is not compatible with the late time observations where we have positive acceleration (note that we have not assume an effective cosmological constant in matter sector i.e. $T_{\mu\nu}$). But the $R=R_0$ solution is the solution which is not sensitive to $\rho_{vac}$'s value and makes the CC natural.
\\	
	
\subsection{A comment on Weinberg's no-go theorem:}
In \cite{weinberg}, Weinberg shows that, under some general assumptions, to cancel the large vacuum density one needs a new fine-tuning (for a short review see \cite{padila}). Though Weinberg's argument is very general but one should be careful to apply it to \"uber-gravity model. In \"uber-gravity we have a very sharp transition from zero to a non-zero value at $R=R_0$ and the Lagrangian is vanishing for $R < R_0$. These properties make \"uber-gravity beyond Weinberg's no-go theorem assumptions. It is  easy to show that in Einstein frame \"uber-gravity in $R<R_0$ will cause all the masses goes to zero\footnote{I would like to thank Justin Khoury for related discussions.} which means our model bypasses Weinberg's argument. Though Weinberg's no-go theorem argues this solution does not describe our real world \cite{padila} but we should emphasize that in \"uber-gravity due to having two different regimes there is a chance to resolve this issue. I.e. $R<R_0$ regime solves the CCP and $R>R_0$ describes the real world. This needs more considerations which remains for future work.

\section{Discussions and concluding remarks:}
Based on the \"uber-modelling idea \cite{uber} we calculated ensemble average of (effectively) all gravity models. This procedure results in an effective Lagrangian (\ref{uberfR1}) which has interesting features shown in FIG.\ref{fig:fR1}. The final Lagrangian (\ref{uberfR1}), \"uber-gravity, is a ``fixed point" in the model space of $f(R)$ models which makes it very special in this space. Below we briefly summarize \"uber-gravity properties:

\begin{itemize}
	\item There is a universal prediction for our model which does not depend on the choice of $\beta$: At high-curvature regime it is the EH gravity which can be crucial to address the local tests. There is always a stronger gravity in an intermediate-curvature regime. For low-curvature regime, $R<R_0$, the gravity is vanishing which can be interpreted as degravitation. Consequently, a sharp transition occurs at $R=R_0$.
	
	\item For any (non-zero) value of $\rho_{vac}$ our model predicts an exact deSitter solution i.e. $R=R_0$. This is an interesting result since  observations support deSitter background. 

	\item If $\rho_{vac}=0$ then our model gives a Minkowski spacetime instead of deSitter spacetime which is not compatible with  observations.

	\item We assume our model works effectively up to Planck mass scale i.e. quantum gravity scale. This means, up to that scale, we do not need to take care of quantum gravity corrections and may conceive our model as a classical field theory. This means even loop corrections to $\rho_{vac}$ cannot change our conclusions and there is no need for infinite regularizing counter terms to keep $R_0$ fixed.
\end{itemize}

According to the above properties, we think \"uber-gravity is a promising model to study in more details. In \cite{Khosravi:2017hfi}, we have introduced a cosmological model, \"u$\Lambda$CDM, based on \"uber-gravity which is a promising solution for $H_0$ tension (e.g. see \cite{Riess:2016jrr}) while it fits background data (including SNe, BAO and first peak of CMB) slightly better than $\Lambda$CDM with very non-trivial predictions at the perturbation level.  There are several ways to pursue this idea which are beyond the scope of this paper and remain open for further investigations:
\renewcommand{\labelitemi}{$\star$}
\begin{itemize}
	\item Making the \"uber-modelling idea more concrete by focusing on its mathematical foundations. Specially we need to address the (fundamental) origin of the probability of each model. 
	\item One specific way to shed light on our model is to calculate effective Newtonian constant. It is doable by a conformal transformation and going from $f(R)$ frame to Brans-Dicke frame.
	\item Doing perturbation theory of our model and examine the results by  observations. This is a crucial test for our model since at the level of background, the solution is exact deSitter which is same as the $\Lambda$CDM.
	\item Focusing on the intermediate-curvature regime which seems potentially attractive. Stronger gravity may address the production of massive black holes in high-redshifts. In addition since the intermediate regime is very close to $R_0$ (which should be fixed by Hubble parameter) then we expect to have features on e.g. CMB in scales very close to Hubble's scale.
	\item We can extend our proposal to other kind of models e.g. Horndeski Lagrangians, massive gravity and other healthy gravity models. 
	\item \"Uber-gravity gives two different phases of gravity depending on the value of Ricci scalar. This suggests that maybe one can think about a phase transition in the cosmology which distinguish early and late time eras.
\end{itemize}

\vskip 0.2 in
\textit{Acknowledgments:}
I would like to thank M. M. Sheikh-Jabbari for very fruitful discussions and useful comments. I am also grateful to S. Baghram, J. Khoury, T. Koivisto, A. Nicolis, A. A. Saberi and S. Shahbazian for useful comments. I also thank anonymous referee for very useful comments which results in improving the paper.  I thank K. Jahan for carefully reading the paper. I also thank School of Physics at IPM for their hospitality during a part of this work.


\begin{thebibliography} {50}
	
	

	
\bibitem{CC-Einstein-1917}A. Einstein, ``Kosmologische Betrachtungen zur allgemeinen Relativitaetstheorie", Sitzungsberichte der Königlich Preussischen Akademie der Wissenschaften Berlin, part 1 (1917) 142–152.	


	



\bibitem{acceleration} Supernova Search Team, ``Observational evidence from supernovae for an accelerating universe and a cosmological constant", Astron.J. 116 (1998) 1009-1038, astro-ph/9805201,\\Supernova Cosmology Project, ``Measurements of Omega and Lambda from 42 high redshift supernovae", Astrophys.J. 517 (1999) 565-586, astro-ph/9812133.  

\bibitem{planck}Planck Collaboration, ``Planck 2015 results. XIII. Cosmological parameters", Astron.Astrophys. 594 (2016) A13, arXiv:1502.01589 [astro-ph.CO].


\bibitem{hobson} M. P. Hobson, G. P. Efstathiou and A. N. Lasenby,``General Relativity: An introduction for physicists" Cambridge University Press, 2006.	


\bibitem{CC-problem} S. Weinberg, ``The cosmological constant problem”, Review of Modern Physics 61 (1989) 1,\\	J. Martin, ``Everything You Always Wanted To Know About The Cosmological Constant Problem (But Were Afraid To Ask)", Comptes Rendus Physique 13 (2012) 566, arXiv:1205.3365 [astro-ph.CO],\\ A. Padilla, ``Lectures on the Cosmological Constant Problem", arXiv:1502.05296 [hep-th].
	
	




\bibitem{CC-MG}N. Kaloper and A. Padilla, ``Sequestering the Standard Model Vacuum Energy", Phys.Rev.Lett. 112 (2014) no.9, 091304, arXiv:1309.6562 [hep-th],\\L. Smolin, ``The Quantization of unimodular gravity and the cosmological constant problems", Phys.Rev. D80 (2009) 084003, arXiv:0904.4841 [hep-th].

\bibitem{CC-FT}N. Arkani-Hamed and S. Dimopoulos, ``Supersymmetric unification without low energy supersymmetry and signatures for fine-tuning at the LHC", JHEP 0506 (2005) 073, hep-th/0405159. 

\bibitem{CC-QG}R. Sundrum, ``Towards an Effective Particle-String Resolution of the Cosmological Constant Problem", JHEP 9907:001 (1999), arXiv:hep-ph/9708329.
	
	
\bibitem{degravitation}	N. Arkani-Hamed, S. Dimopoulos, G. Dvali and G. Gabadadze, ``Nonlocal modification of gravity and the cosmological constant problem", hep-th/0209227,\\ G. Dvali, S. Hofmann and J. Khoury, ``Degravitation of the cosmological constant and graviton width", Phys.Rev. D76 (2007) 084006, hep-th/0703027.  


	
	

	
	
\bibitem{supersym}S. P. Martin,``A Supersymmetry primer",  Adv.Ser.Direct.High Energy Phys. 21 (2010) 1-153, hep-ph/9709356.

	
	
\bibitem{uber}N. Khosravi, ``Ensemble Average Theory of Gravity", Phys.Rev. D94 (2016) no.12, 124035, arXiv:1606.01887 [gr-qc].		


\bibitem{nnaturalness} N. Arkani-Hamed, T. Cohen, R. T. D'Agnolo, A. Hook, H. D. Kim and D. Pinner, ``Solving the Hierarchy Problem at Reheating with a Large Number of Degrees of Freedom", Phys.Rev.Lett. 117 (2016) no.25, 251801,  arXiv:1607.06821 [hep-ph].


\bibitem{tegmark} M. Tegmark, ``Is "the theory of everything'' merely the ultimate ensemble theory?", Annals Phys. 270 (1998) 1-51, arXiv:gr-qc/9704009\\
M. Tegmark, ``The Mathematical Universe", Found. Phys. 38 (2008) 101-150, arXiv:0704.0646 [gr-qc].



\bibitem{Khosravi:2017hfi} 
N.~Khosravi, S.~Baghram, N.~Afshordi and N.~Altamirano,
``\"u$\Lambda$CDM: $H_0$ tension as a hint for \"Uber-Gravity,''
arXiv:1710.09366 [astro-ph.CO].

\bibitem{fR-review}A. De Felice and S. Tsujikawa, ``f(R) theories", Living Rev. Rel. 13: 3, 2010, 	arXiv:1002.4928 [gr-qc].

\bibitem{Rn}S. Carloni, P. K. S. Dunsby, S. Capozziello and A. Troisi, ``Cosmological dynamics of $R^n$ gravity", Class. Quant. Grav. 22 (2005) 4839-4868,  gr-qc/0410046.



\bibitem{axion}D. J. E. Marsh, ``Axion Cosmology", Phys.Rept. 643 (2016) 1-79, arXiv:1510.07633 [astro-ph.CO],\\
P. Svrcek and E. Witten, ``Axions In String Theory", JHEP 0606 (2006) 051, hep-th/0605206.


\bibitem{witten}L. Hui, J. P. Ostriker, S. Tremaine and E. Witten, ``Ultralight scalars as cosmological dark matter", Phys.Rev. D95 (2017) no.4, 043541, arXiv:1610.08297 [astro-ph.CO].



\bibitem{weinberg}
S.~Weinberg,
``The Cosmological Constant Problem,''
Rev.\ Mod.\ Phys.\  {\bf 61}, 1 (1989).

\bibitem{padila}
A.~Padilla,
``Lectures on the Cosmological Constant Problem,''
arXiv:1502.05296 [hep-th].


\bibitem{Riess:2016jrr}
A.~G.~Riess {\it et al.},
``A 2.4$\%$ Determination of the Local Value of the Hubble Constant,''
Astrophys.\ J.\  {\bf 826}, no. 1, 56 (2016)
[arXiv:1604.01424 [astro-ph.CO]].





\end{thebibliography}
\end{document}